# Photo-response of the conductivity in functionalized pentacene compounds

T. Tokumoto[1*], J.S. Brooks[1], R. Clinite[2], X. Wei[1], J.E. Anthony[3], D.L. Eaton[3], S. R. Parkin[3]

Physics Dept. and NHMFL, Florida State University, Tallahassee FL 32310 USA
Physics Dept., Cornell University, Ithaca, NY, 14850 USA
Department of Chemistry, University of Kentucky, Lexington, KY 40506 USA

Abstract

   We report the first investigation of the photo-response of the conductivity of a new class of organic semiconductors based on functionalized pentacene. These materials form high quality single crystals that exhibit a thermally activated resistivity. Unlike pure pentacene, the functionalized derivatives are readily soluble in acetone, and can be evaporated or spin-cast as thin films for potential device applications. The electrical conductivity of the single crystal materials is noticeably sensitive to ambient light changes. The purpose, therefore, of the present study, is to determine the nature of the photo-response in terms of carrier activation vs. heating effects, and also to measure the dependence of the photo-response on photon energy. We describe a new method, involving the temperature dependent photo-response, which allows an unambiguous identification of the signature of heating effects in materials with a thermally activated conductivity. We find strong evidence that the photo-response in the materials investigated is predominantly a highly localized heating mechanism. Wavelength dependent studies of the photo-response reveal resonant features and cut-offs that indicate the photon energy absorption is related to the electronic structure of the material.

Corresponding author: brooks@magnet.fsu.edu

Draft date:        Monday, March 25, 2002

I. INTRODUCTION

The synthesis of a novel class of pentacene-based materials with improved interplanar molecular stacking has recently been reported[1]. Temperature dependent resistivity studies[2] have shown that these materials are semiconductors, with thermal activation energies in the range of 0.3 to 1.5 eV. Recent band structure calculations[3] are consistent with the semiconductor description in terms of experimental values of the energy gaps, and their observed anisotropy. The semiconductor character of these materials, combined with their ease of synthesis and processing, makes them promising candidates for application as sensors and devices, and hence their full physical characterization is important. In the present paper, we discuss the photo-response of the conductivity of two of these materials, in the form of single crystals. Here triisopropylsilylethynyl (TIPS) or trimethylsilylethynyl (TMS) side groups are attached to either side of the central ring sites (position 6 and 13) of the pentacene structure, as shown in Fig. 1. The main structural difference between the two materials is in the π – stacking of the acene arrays, which for TIPS is two-dimensional, and for TMS is one-dimensional. In the study presented here, for both materials un-polarized light is incident in the plane normal to the molecular stacking direction.

Of note is that unlike crystalline pentacene, the functionalized pentacene derivatives are very soluble, and single crystals of several $mm^2$ in area and 50 μm in thickness are readily precipitated from acetone. These derivatives are also significantly (approximately two times) more oxidatively stable than pure pentacene, and the electrical and photo-dependent characteristics of individual crystals are stable for periods of a year or more under ambient light and atmospheric conditions. Single crystals appear dark blue in color, mainly due to reflected light. Under transmitted light, a reddish color is sometimes observed *due to the efficient fluorescence emission of the pentacene moiety*. As shown below, this effect is measurable in the wavelength dependence of the photo-response for a particular contact lead configuration. *Here we define the photo-response as the increase in current in a sample for a constant current or voltage bias across the two-terminal contacts made to the sample* (see Fig. 2). The resistance of the samples is very sensitive to ambient light conditions (of order 5 to 10 % for florescent (or incandescent) room lights turned on and off). By investigating carefully the temperature dependence of the resistance and the photo-response, we have determined that the photo-response is primarily due to the absorption and rapid thermalization of light, which in turn heats the sample and reduces the temperature dependent resistance at the surface where the transport currents are concentrated. These materials could, at the present stage of development, be used as a prototype pixelated photo-detector that could discriminate between red and blue light.

II. EXPERIMENT

Single crystals of the materials under investigation were synthesized by methods that have been described elsewhere[1]. Electrical contact was made to the samples with 25 micron gold wires and silver or graphite paint. Conventional constant current (or voltage) sources were used, and the dc sample voltage was monitored with an electrometer with a 200 T$\Omega$ input impedance. The room temperature resistance of a typical sample is of order 0.1 to 100 G $\Omega$, with the contact resistances of order 1/10 of the sample resistance. Here we note that silver paint yields significantly better electrical contact than the graphite paint. The experimental setup for the photo-response studies is shown in Fig. 2. For the temperature dependent measurements, a vacuum chamber was used to avoid any interaction with moisture etc., and the light excitation was provided by a standard 0.5 mW He-Ne laser (632.8 nm) chopped at about 100 Hz. The wavelength dependence of the photo-response was carried out at room temperature in air. Here a halogen light source was used with a spectral grating (McPherson spectrometer system) to scan the wavelength from 350 to 1400 nm. In the range 700 to 1400 nm, a cutoff filter ($\lambda_{cutoff} <$ 700 nm) was used to attenuate any secondary reflections from lower wavelength light.

The photo-response ($V_{ac}$) and sample bias voltage ($V_{dc}$) were measured as shown in Fig. 2. Due to the very high resistance of the samples when compared with the contact resistance, the sample properties dominate the behavior of the measurements. (We note that previous measurements to obtain the temperature and crystallographic orientation dependent resistivity were done in a four terminal configuration.) Measurements were carried out either in a constant current or constant voltage configuration. Changes in the current through the sample, in response to the chopped light excitation at about 100 Hz, were detected by the ac voltage ($V_{ac}$) across a 1 M$\Omega$ resistor in series with the bias circuit. A lock-in amplifier was used to measure the ac signal, which was of order 1 mV in our experiment. In the configuration in Fig. 2, due to significant RC time constants, we can only estimate a lower bound on the response time of the photo-response, which we found to be less than 1 microsecond from oscilloscope traces.

III. MECHANISMS OF PHOTO-SENSITIVE CONDUCTIVITY

An important issue question concerning semiconductors that show an increase in the conductivity due to light excitation is the origin of the mechanism, i.e., heating and/or carrier excitation. Both of these effects will lead to the same result – a decrease of resistance that for a semiconductor has an exponential dependence on temperature. Therefore, to address this question in the present case, we have analyzed in detail the photoconductivity and heating mechanisms as they are manifested in our measurement scheme.

We first consider the case of pure excitation of carriers (photoconductivity). Since the changes in the bias current $I_0$ and voltage $V_{dc}$ are very large with respect to the ac variation in current $\Delta I$ (over the period $\Delta t$), and voltage $V_{ac}$ (= $\Delta V_X/\Delta t$) we may assume an adiabatic limit where the bias circuit does not respond to the small ac changes. Also, the current resistor X is negligible compared with the sample resistance. Hence we may relate the change in current to the change in sample resistance as $\Delta I = -\Delta R\, I/R$. Likewise, $V_{ac} = X\, \Delta I = -X\, \Delta R\, I/R = -X\, \Delta R\, IR/R^2 = X\, G V_{dc}$, where the photoconductivity G is defined as $G = \Delta(1/R) = -\Delta R/R^2$. Hence $V_{ac}$, in the case of excited carriers, will be proportional to both the photoconductivity G and the bias potential $V_{dc}$. Finally then, $G = V_{ac}/X V_{dc}$.

We may now consider the case of pure heating. We assume a simple model for the sample in terms of its activated resistance $R = R_0 \exp(E_a/kT)$ where $E_a$ is the thermal activation energy, and its lattice heat capacity $C = C_0 T^3$ which we take as purely phonon dominated. We further assume that the laser heat pulse ($\Delta Q$) does not change with temperature, so that the change in the temperature of the sample with a pulse is $\Delta T = \Delta Q/C_0 T^3$. When the sample temperature changes by $\Delta T$, the resistance of the sample will change by $\Delta R = dR/dT\, \Delta T = -R_0 E_a/kT^2 \exp(E_a/kT)\, \Delta T = -R_0 E_a/kT^2 \exp(E_a/kT)\, \Delta Q/C_0 T^3$. Hence the ac response due to heating has the form $V_{ac} = X\, \Delta I = -X\, \Delta R\, I/R = X\, R_0 E_a/kT^2 \exp(E_a/kT)\, \Delta Q/C_0 T^3\, I/R$. It is easy to show that the temperature dependence of $V_{ac}$ has two possible forms:

$V_{ac} \sim 1/T^5$ for a constant current bias  (1)

$V_{ac} \sim \exp(-E_a/kT)/T^5$ for a constant voltage bias  (2)

Hence for a constant current bias, the photo-response will decrease with increasing temperature, whereas for a constant voltage bias it will increase. One further prediction one can make is to consider data in the form $V_{ac}$ vs. I at constant temperature (and average sample resistance). From the above, the slope will be proportional to $1/T^5$. These relations provide a rigorous method for determining the role of heating in photo-sensitive materials.

IV. RESULTS OF TEMPERATURE DEPENDENT MEASUREMENTS

Shown in Figs. 3 and 4 are typical results for the temperature dependence of the photo-response of the TIPS and TMS samples, expressed in terms of the dc bias voltage and ac photo-response. Experimental values for $V_{ac}$ were generally in the mV range (across a 1 MΩ resistor), $V_{dc}$ was limited to 100 V maximum, the corresponding currents used were in the range of 1 to 100 nA, and sample resistances ranged from 100 MΩ to 100 TΩ. The temperature of the sample was increased using a heater platform placed in a vacuum, where at maximum temperature, the platform was allowed to relax back to room temperature over several hours. We see that for constant current bias, the sample resistance (and voltage bias) decrease with increasing temperature according to a thermal activation law (see also Ref. 2), but <u>the temperature dependence of the ac photo-response depends on the sample bias condition,</u> i.e. for constant voltage bias the photo-response increases with increasing temperature. Indeed, the bias and temperature dependence of the photo-response follows closely the predictions given above in Eqs. 1 and 2. In Fig. 3c we show measurements that test the prediction that the slopes of the $V_{ac}$ vs. I curves should follow an inverse power law in temperature and this indeed is the case.

Collection of the temperature-dependent TMS data in Fig. 4 was motivated in part by preliminary temperature dependent X-ray structural data[4]. Here, a significant collective vibration along the long-axis of the pentacene moiety appears to become the dominant thermal motion. Furthermore, a monotonic increase in the distance between pentacene units along the _-stacking axis was found, with the spacing changing from an average 3.44 Å at -100 °C to 3.51 Å at 57 °C. This distance did *not* increase further upon increasing temperature. Correspondingly, we found that in the vicinity of 50 °C the resistance of the sample became nearly temperature independent. This behavior is consistent with an increase in the intrinsic resistivity with temperature, which competes with the thermal activation process. The increase in resistivity may be the result of a less favorable π-overlap that is related to the carbon site elongation, but this possibility must await a full crystallographic X-ray analysis. The data in Fig. 4a was carried out for a constant current bias, but the sample was not monitored after reaching room temperature. In Fig. 4b , the same sample was studied a day later for constant voltage bias, and this time data was taken for several hours after room temperature was recovered. Here the sample was observed to relax back to its initial state with time, and the temperature dependent structural changes therefore appear to be reversible.

V. WAVELENGTH DEPENDENCE OF THE PHOTO-RESPONSE

The temperature dependence of the photo-response discussed above was evaluated at a fixed wavelength of 632.8 nm. In this section we describe the wavelength dependence of the photo response from 350 to 1100 nm at room temperature. Since the light source was thermal (a halogen lamp with a McPherson grating analyzer), we show in Fig. 5 the estimated wavelength dependence of the light intensity in terms of the theoretical black body spectrum, and also the experimental response of a silicon detector. The data were obtained without a filter in the range 350 ~ 800 nm, and with a filter ($\lambda_{cutoff}$<700 nm) in the range 700 ~ 1400 nm. In Figure 6 the photo-response of the TIPS and TMS samples are shown. We observed no photo-response for wavelengths in the range 1100 to 1400 nm, even though the light intensity was still significant in this range, as Fig. 5 indicates. Light was incident on the sample in the plane normal to the $\pi$-orbital stacking direction, either from the contact (forward), or non-contact (back) sides as shown in Figure 2. The most salient features of the data are: 1) the onset of photo-response below about 950 nm; 2) the strong attenuation of the photo-response for the back-illumination data below about 750 nm; and 3) the wavelength-dependent structure for the case of forward illumination.

VI. Discussion.

We believe that the primary mechanism of photo-induced conductivity as measured in the present technique is one where light energy is absorbed, thereby heating the sample locally in the region where the current density is greatest between the electrical contacts. This conclusion is based on temperature dependent and wavelength dependent photo-response. Furthermore, the onset of adsorption of light appears to be related to the solid-state energy band edge, and to other, higher energy (intra-molecular) transitions in the materials studied. We find that a simple model for the photo-response, based on the thermally activated conductivity, gives an accurate description of the data. The advantage of this model is that it predicts two distinctly different forms of the temperature dependence, depending on the two (constant current or voltage) bias conditions.

That the heating effect is highly localized is made further evident from the wavelength dependent studies where the sample contacts are made on one side of the sample. Here we find that for longer wavelengths below the solid-state band edge for the material ($\lambda_{band}$ ~ 800 nm), the sample is more transparent and light can reach the contact region where there is a high current density. (The solid-state band edge is estimated from recent band structure calculations by Haddon et al.[3], which range from about 0.5 (2480 nm) to 1.5 eV (825 nm) over the three crystallographic directions in these materials.) However, at shorter wavelengths above the band edge, the light energy is absorbed and does not reach the contact region, and hence no photo-response is observed.

For light incident on the contact side, the photo-response amplitude generally follows the black-body spectrum, but additional structure is observed that indicates that the energy adsorbed is sensitive to the electronic structure of the material. We have made a comparison between the photo-response and UV-Vis data for both materials as is shown in Fig. 7. Here we have normalized the photo-response intensity by using the photodiode calibration from Fig. 5. The normalized photo-response shows, in both samples, a pronounced feature near 410 nm, which we believe to be associated with the $L_B$ dipolar intra-molecular mode along the long pentacene molecular axis. Notably, this seems most prominent in the TMS system, which shows a reversible thermal transition in the vicinity of 50 °C (See Figure 4). Recent temperature dependent X-ray refinement data[4] at 57 °C also show an increase in carbon site displacements in this same direction. We may speculate at this point that the unusual behavior of the $L_B$ mode is responsible for its prominence in both the spectral- and temperature-dependent features in the photo-response of the TMS system. Although other spectral (resonant) features clearly appear in the photo-response in the range 400 to 800 nm, they are more difficult to correlate directly with specific molecular vibrational spectra. The photo-response of the TIPS material most closely resembles the UV-Vis spectrum in this range.

The difference between front and back illumination was studied for both silver paint and graphite paint contacts. This was done to examine the possible influence between chemical or work-function (photodiode) related properties[5] of the contacts. Although the graphite paint contacts were generally higher in resistance, compared with silver paint contacts, with a reduced signal-to-noise ratio, there was no clear qualitative difference in the photo-response in the two cases. The temperature dependence of the photo-response rules out a metal-semiconductor interface photodiode behavior, which should not change with bias condition – that is, a photodiode current should be thermally assisted, and increase regardless of the bias condition.

We have made a simple estimate of the quantum efficiency (which is an indication of the number of photo-excited carriers) and the temperature rise in the sample due to conventional thermalized photon energy. The rms photo-induced current through the sample was typically of order 1 mV/MΩ per cycle, or $1 \times 10^{-9}$ Coul/sec/cycle, which for about 100 Hz is $1 \times 10^{-11}$ C/pulse. The power of the laser was 0.5 mW, which, for 600 nm light is $5 \times 10^{14}$ Hz, so the quantum efficiency, assuming full adsorption of the laser light, is: QE = $(1 \times 10^{-11} C/e)/(5 \times 10^{-4} W * .01 s/h * 5 \times 10^{14} Hz) = 6.24 \times 10^{7}$ electrons/$1.5 \times 10^{13}$ photons = $4 \times 10^{-6}$ electrons/photons. From this we find that the carrier excitation process appears to be negligible, even if we take into account the fact that perhaps 80% of the laser beam is reflected. Likewise, the heating effect may be estimated from the temperature rise due to the laser pulse. Here we consider that the specific heat is c= 0.1 J/g°C, and the laser pulse deposits a heat of $5 \times 10^{-6}$ J per pulse. For a typical sample of 1mm$^2$ x 50 microns, and a density approximately that of water, we find that C =mc = $50 \times 10^{-6}$ g *0.1J/g°C. So, $\Delta T = \Delta Q/C = 5 \times 10^{-6}/5 \times 10^{-6} = 1$ °C. A change of 1 °C would be sufficient to reduce the equilibrium resistance by several percent, and hence this could easily cause the observed increase in current. We note however that photo-response relaxation time was very fast, and was not observable on the millisecond scale of our measurements. Therefore if heating plays a role, it is confined to a very small region of the sample where the current density is a maximum in the vicinity of the contacts.

VII. SUMMARY

Funtionalized pentacene compounds in the solid state hold promise for potential applications in photonic devices due to the high sensitivity (and relatively high speed) of their resistive properties, in response to light, coupled with their wavelength selectivity due to adsorption processes related to their electronic structure. The ease of synthesis and processing increases their potential usefulness in device applications. In the present investigation of these materials, we have developed and employed a method to discriminate between optical thermalization and a photo-carrier excitation processes in materials with activated conductivities. Future work will focus on the electronic and photosensitive thin film devices. Here it has been found that for sublimation, the pentacene moieties order crystallographically on certain subtrates[6].

ACKNOWLEGEMENTS

The FSU group is supported through NSF-DMR 9971474, and one of us (RC) is grateful for a REU-NSF fellowship during the course of this work. The UK group acknowledges support from the Office of Naval Research (N00014-99-1-0859) and a Kentucky Opportunity Fellowship (DLE).

FIGURE CAPTIONS

Figure 1. Molecular structure for the TIPS and TMS materials. In both cases, the (un-polarized) light is incident in the plane normal to the $\pi$-orbital stacking direction.

Figure 2. Experimental setup for the photo-response measurements. Current is supplied to the sample ( R ) either in a constant current, or constant voltage mode. The 1 M$\Omega$ input impedance (X) of an oscilloscope serves as the current monitoring series resistor. The dc bias of the series circuit is measured with a high impedance (200 T$\Omega$) electrometer, and the ac signal is detected with a phase sensitive detector (lock-in amplifier). The circuit analysis is described in the text. Light from either the laser or the spectrometer is incident on the sample surface, either from the contact, or non-contact sides. The sample temperature may be varied in vacuum from room temperature to 100 °C by application of heat to the sample-mounting platform.

Figure 3. The temperature dependence of the photo-response of a TIPS sample in vacuum. a) $V_{dc}$ and $V_{ac}$ (photo-response) signals for a constant current bias. b) $V_{ac}$ signal for constant dc bias (100 V). c) $V_{ac}$(photo-response) vs. current for different temperatures. Inset: temperature dependence of the slope of $V_{ac}$ vs. current data. Solid line, theoretical $1/T^5$ dependence. (See text for discussion of temperature and bias dependence.)

Figure 4. The temperature dependence of the photo-response response of a TMS sample in vacuum. a) ) $V_{dc}$ and $V_{ac}$ (photo-response) signals for a constant current bias. b ) $V_{ac}$ (photo-response) signal for constant dc bias (100 V). Here data was taken for several hours after the warming and cooling cycle to monitor the relaxation of the TMS structural transition that occurs near 50 to 60 degrees. (See text for discussion.)

Figure 5. Black body spectrum (theoretical) and the non-normalized response of silicon photodiode (measured) for the halogen light source and spectrometer output. Optical scans were made without a cutoff filter from 300 to 800 nm, and with a cutoff filter from 700 to 1400 nm.

Figure 6. Photo-response for forward and backward illumination for a) TIPS sample with silver paint electrodes on front side, b) TIPS sample with graphite paint electrodes on front side, c) TMS sample with silver paint contacts that extended to both sides. All data were taken at overlapping wavelengths using a cutoff filter for the 800 to 1100 nm range.

Figure 7. Comparison of UV-Vis spectrum with the forward illumination photo-response data in the 350 to 800 nm range normalized with the (structureless) wavelength dependent intensity measured with the silicon photodiode ( from Fig. 5). a) TIPS sample. b) TMS sample.

- 

FIGURES

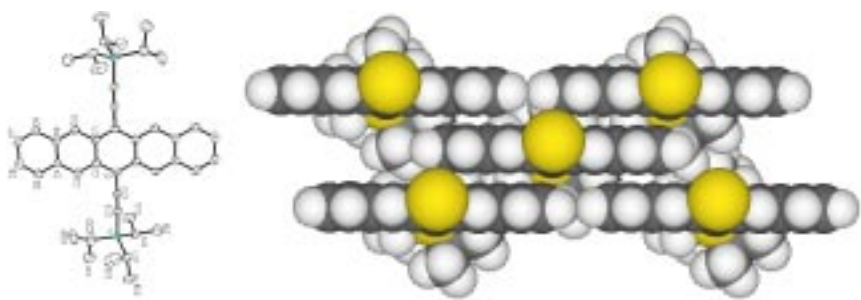

**Figure 1a.** Triisopropylsilyl (TIPS) derivative: Structure and solid-state order (minus alkyl groups).

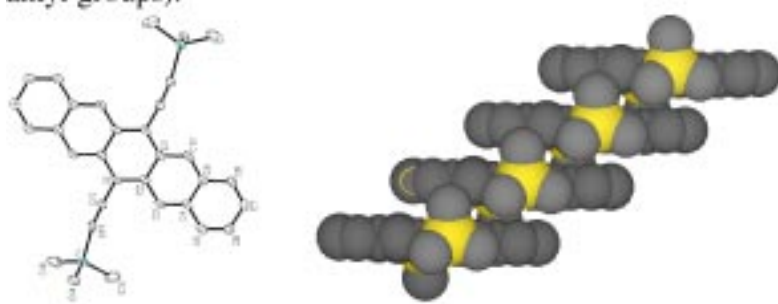

**Figure 1b.** Trimethylsilyl (TMS) derivative: Structure and solid-state order.

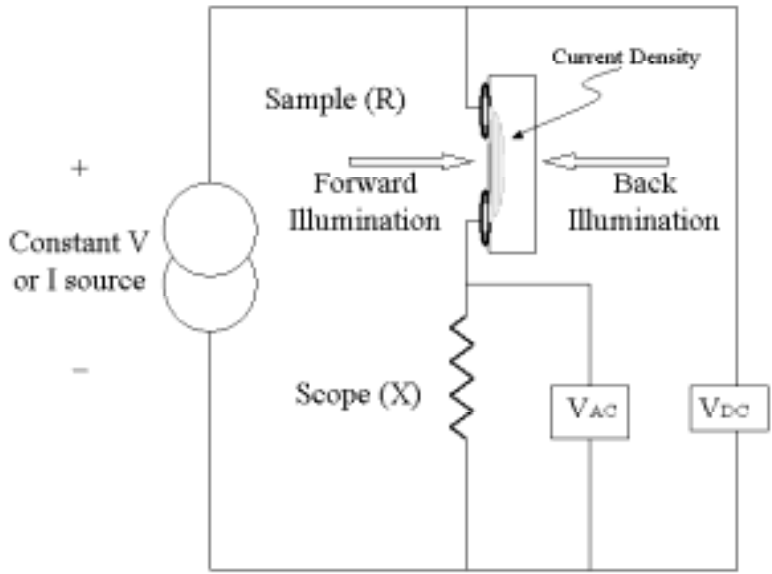

**Figure 2**

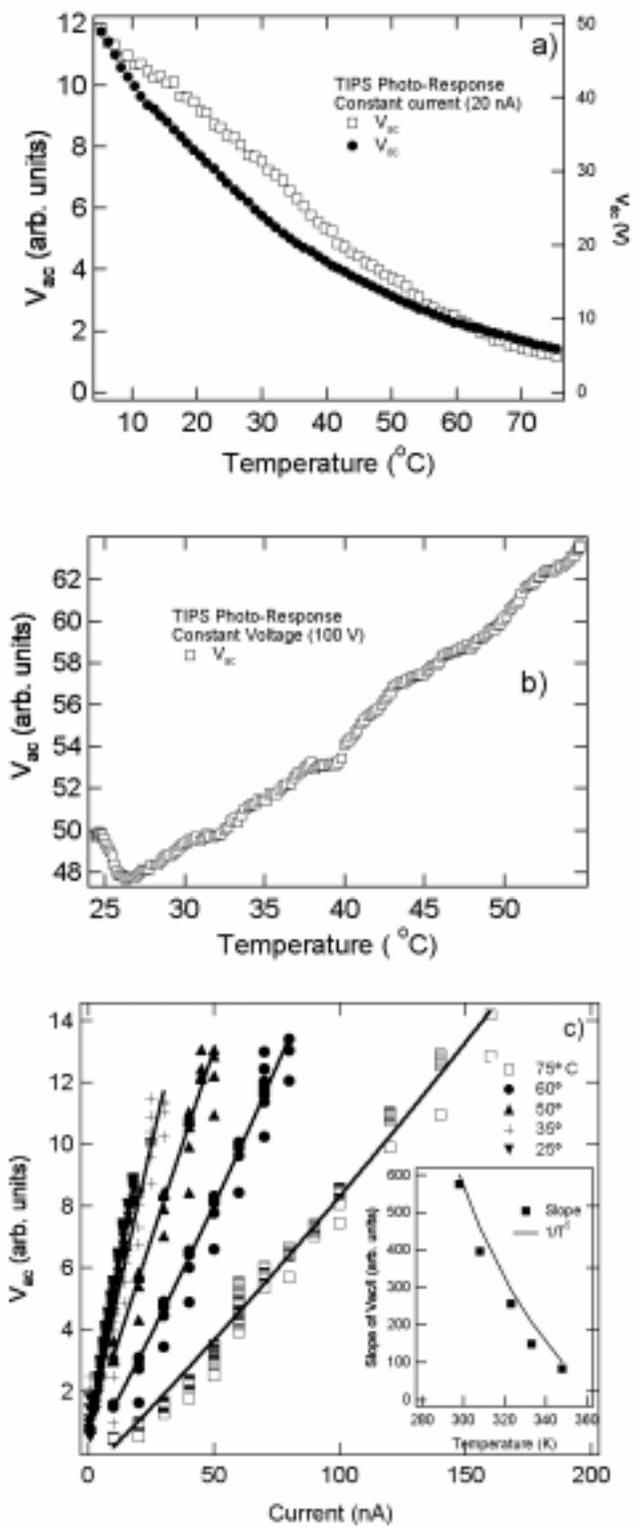

**Figure 3 a,b,c**

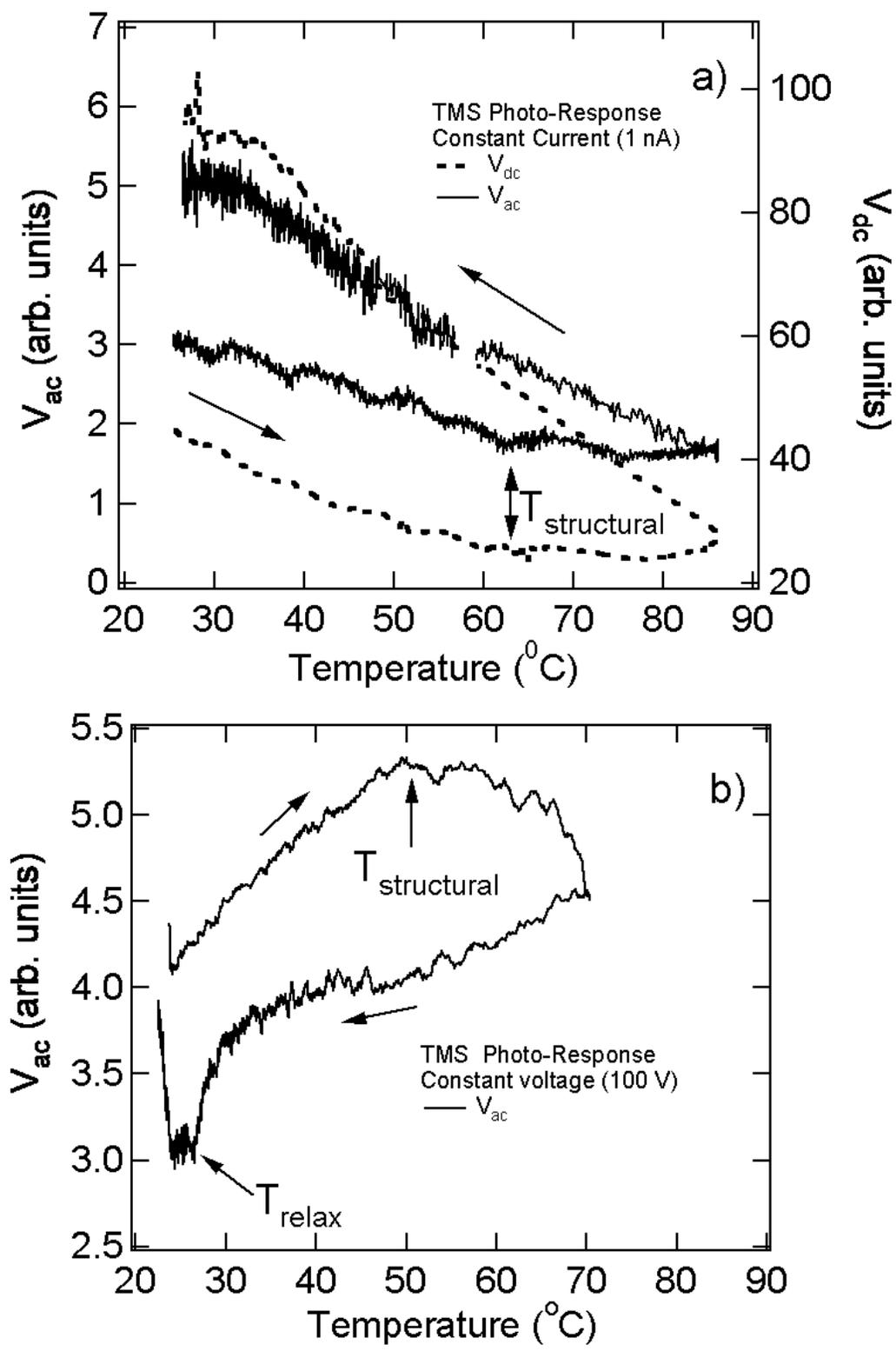

**Figure 4 a,b**

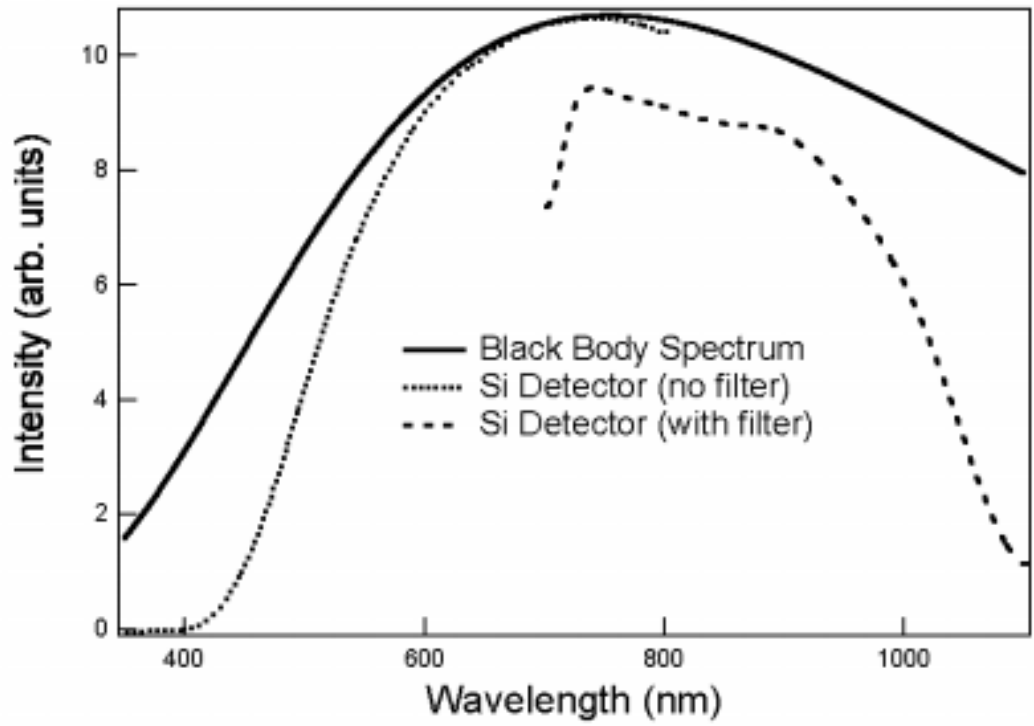

**Figure 5**

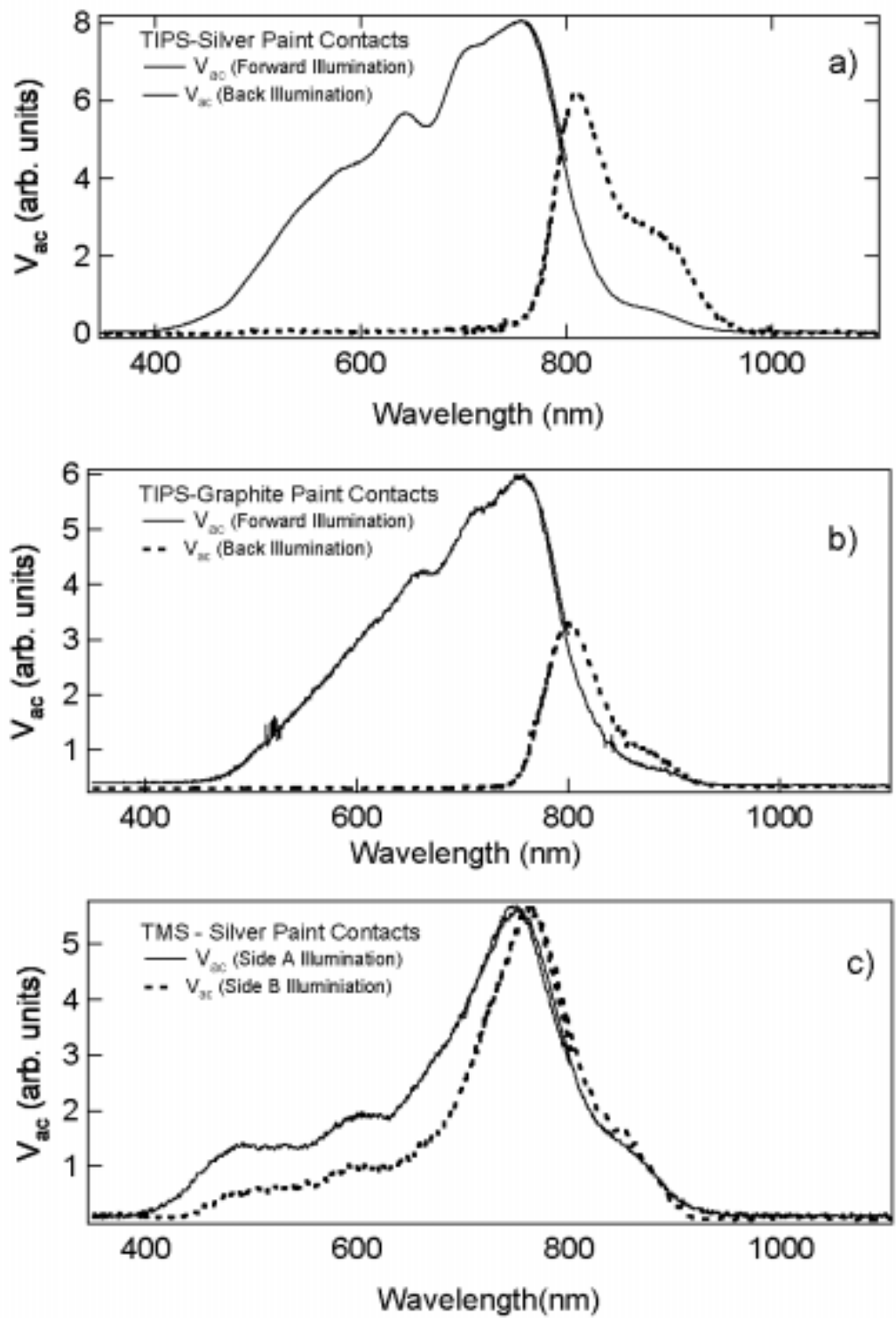

**Figure 6 a,b,c**

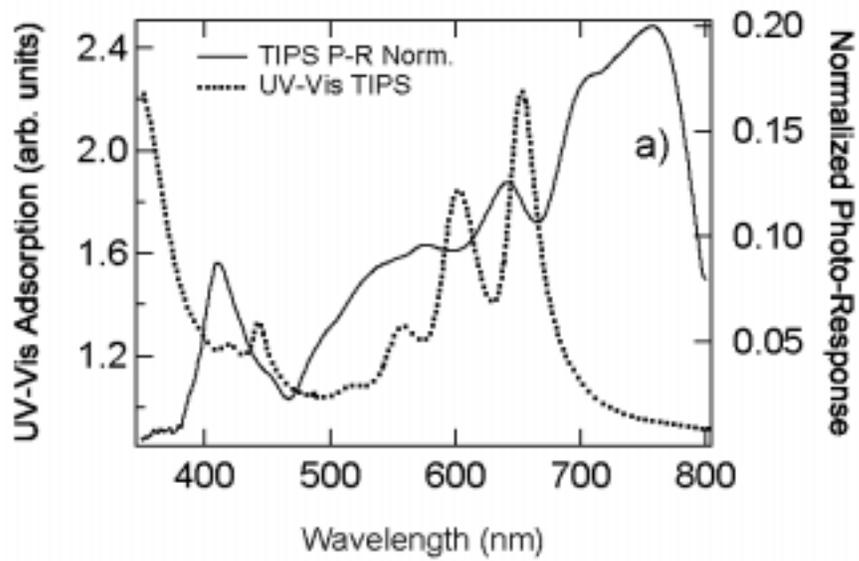

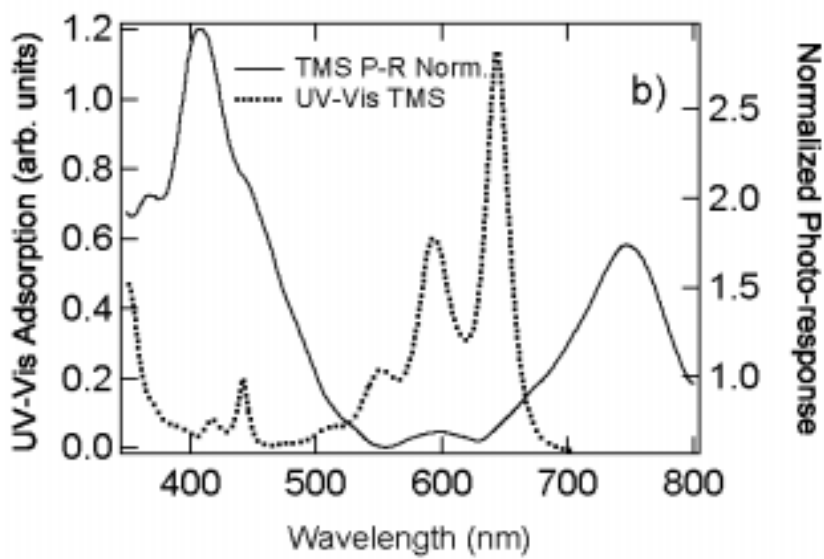

**Figure 7 a,b**